# Performance Analysis of Apriori Algorithm with Different Data Structures on Hadoop Cluster

Sudhakar Singh
Dept. of Computer Science
Faculty of Science
Banaras Hindu University

Rakhi Garg
Dept. of Computer Science
Mahila Maha Vidyalaya
Banaras Hindu University

P.K. Mishra
Dept. of Computer Science
Faculty of Science
Banaras Hindu University

## ABSTRACT
Mining frequent itemsets from massive datasets is always being a most important problem of data mining. Apriori is the most popular and simplest algorithm for frequent itemset mining. To enhance the efficiency and scalability of Apriori, a number of algorithms have been proposed addressing the design of efficient data structures, minimizing database scan and parallel and distributed processing. MapReduce is the emerging parallel and distributed technology to process big datasets on Hadoop Cluster. To mine big datasets it is essential to re-design the data mining algorithm on this new paradigm. In this paper, we implement three variations of Apriori algorithm using data structures hash tree, trie and hash table trie i.e. trie with hash technique on MapReduce paradigm. We emphasize and investigate the significance of these three data structures for Apriori algorithm on Hadoop cluster, which has not been given attention yet. Experiments are carried out on both real life and synthetic datasets which shows that hash table trie data structures performs far better than trie and hash tree in terms of execution time. Moreover the performance in case of hash tree becomes worst.

## 1. INTRODUCTION
We are surrounded with excessive amount of digital data but ravenous for potentially precise information. Data mining is the technique that finds hidden insight and unknown patterns from massive database, which are used as useful knowledge for decision making. Association Rule Mining (ARM) [1] is one of the most important functionality of data mining that comprises of two tasks; finding frequent itemsets and finding interesting correlation between set of frequent items. An itemset is said to be frequent if its support is greater than or equal to a user defined minimum support threshold. Support of an itemset is the percentage of transactions containing that itemset in database. The Apriori algorithm proposed by R. Agrawal and R. Srikant [2] is the most widely used algorithm for mining frequent itemset. Various data structures and a number of sequential and parallel algorithms have been designed to enhance the performance of Apriori algorithm.

Big Data [3] technologies create a biggest hype just after its emergence. A new parallel and distributed computing paradigm has been introduced which is largely scalable and does not require high-end machines. Hadoop is such a large-scale distributed batch processing infrastructure for parallel processing of big data on large cluster of commodity computers [4]. MapReduce is an efficient and scalable parallel programming model of Hadoop that process large volumes of data in parallel and distributed fashion. Traditional tools and techniques of data mining are not scalable and efficient to manage big data. Recent advances are porting data mining algorithms on this new paradigm. Many authors have re-designed and implemented the Apriori algorithm on MapReduce framework in an efficient way but the impact of data structures on the efficiency of MapReduce based Apriori algorithm have not been yet evaluated.

Data structures are the integral in designing of any algorithm. A well-organized data structure significantly reduces the time and space complexity. Apriori algorithm finds the frequent itemsets by generating a large number of candidate itemsets. Candidates are the itemsets containing all potentially frequent itemsets. To make candidate generation efficient and to optimize space for storing intermediate candidates various data structures have been designed by many authors; among them most eminent are Hash Tree, Trie (Prefix Tree) and Hash Table Trie [2] [5-6]. In the sequential implementation of Apriori, trie performs better than hash tree [5] but hash table trie does not perform faster than trie [6]. In this paper, we describe the implementations and evaluate the Apriori algorithms based on three data structures in MapReduce context. Experimental results on both real life and synthetic datasets show that hash table trie takes very less execution time as compared to trie. Also the execution times of trie and hash tree are of the same order as it was in sequential Apriori.

The rest of the paper is organized as follows. Section 2 describes the central data structures used in Apriori algorithm and also introduces the Hadoop system. Related works are summarized in section 3. Section 4 gives the implementation details of Apriori on MapReduce framework. Experimental results are evaluated in section 5. Finally section 6 concludes the paper.

## 2. BACKGROUND
In this section we briefly describe the Apriori algorithm and a comparative overview of hash tree and trie data structures. We also discuss the MapReduce programming paradigm and Hadoop Distributed File System (HDFS) of Hadoop.

### 2.1 Apriori Algorithm
Apriori is an iterative algorithm which generates frequent 1-itemsets $L_1$ by scanning the whole database in first iteration. In k$^{th}$ iteration ($k \geq 2$) it generates candidate k-itemsets $C_k$ from frequent (k-1)-itemsets $L_{k-1}$ of last iteration. Again whole database is scanned to count the support of candidate itemsets by checking subset of each transaction to be candidate. Candidates having minimum support are resulted as frequent k-itemsets $L_k$. Generation of candidates $C_k$ from frequent itemsets $L_{k-1}$ consists of two steps join and prune. In join step, $L_{k-1}$ is joined with itself with condition that two itemsets of $L_{k-1}$ are joined if their first *(k-2)* items are same and *(k-1)*$^{th}$ item of first itemset is lexicographically less than the respective item of second itemset. Prune step reduces the size of $C_k$ using Apriori property. Apriori property states that any (k-1)-itemset that is not frequent cannot be a subset of a frequent k-itemset [2] [7].Joining and pruning of itemsets and checking subset of each transaction against candidates are very computation intensive process in Apriori algorithm. Also a large number of candidates require large memory during execution of algorithm. Therefore, an efficient data structure is required, which reduces the computation cost as well as organizes candidate itemsets in a compact way in memory. Hash Tree and Trie are the central data structure gratifying this requirement. In the next subsection we compare the three data structure in the context of their operations in Apriori algorithm.

## 2.2 Hash Tree vs. Trie

Hash tree and trie both are rooted (downward), directed tree. Hash tree contains two types of nodes, inner nodes and leaves. Leaves of hash tree contain a list which stores candidates. Every inner node stores a hash-table which directs to the nodes at next level downward applying a hash function. At the leaf nodes if the number of candidates exceeds a threshold value *leaf_max_size*, then leaf node is converted to an inner node. Trie does not differentiate between its inner node and leaves. It stores an itemset on a path form root to a particular node. There are links between nodes of two consecutive levels and each link is associated with an item. It requires less memory to store candidates since common prefixes are stored only once [5].

Let a set of items *I = {i1, i2, i3, i4, i5}*. Suppose all 3-itemsets generated from *I* as candidates, then $C_3$ *= {i1 i2 i3; i1 i2 i4; i1 i2 i5; i1 i3 i4; i1 i3 i5; i1 i4 i5; i2 i3 i4; i2 i3 i5; i2 i4 i5; i3 i4 i5}*. Let *child_max_size* (maximum number of child nodes or table size) to be 3 and a hash function defined over items as *h (item) = item % child_max_size*. All items are assigned a corresponding numerical value so that hash function can be applied. Now the set of items *I* and candidates $C_3$ can be represented as *I = {1, 2, 3, 4, 5}*and $C_3$ *= {1 2 3; 1 2 4; 1 2 5; 1 3 4; 1 3 5; 1 4 5; 2 3 4; 2 3 5; 2 4 5; 3 4 5}*. Figure 1(a) and (b) shows the hash tree and trie containing same number of candidates $C_3$, represented differently.

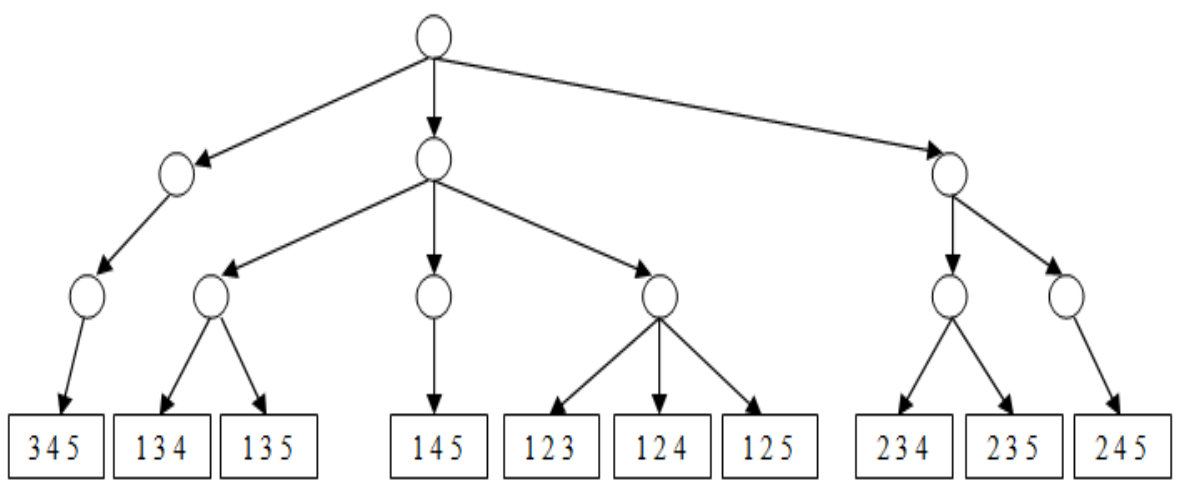

**Fig 1(a): A Hash Tree with 10 candidates**

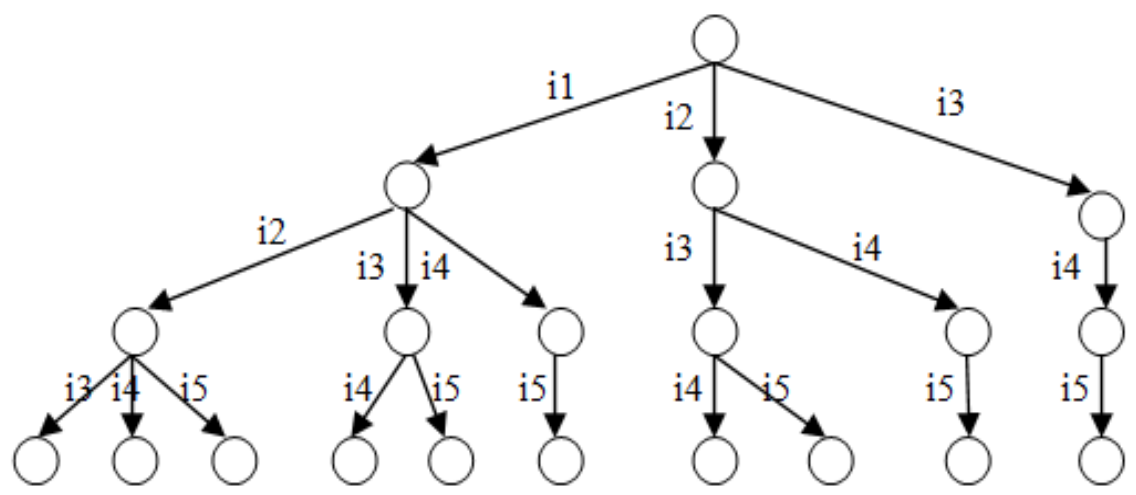

**Fig 1(b): A Trie with 10 candidates**

Theoretically trie is faster and candidate generation is simple in comparison to hash tree. Hash tree is slow in retrieval operation due to two phases of operation. Particularly for the support counting, one first has to traverse to the leaf node and then search in the list of candidates at leaf node whereas in trie only need to traverse to the leaf node. Candidate generation is simple in trie since it is easy to find common (k-1)-prefix to generate candidate k-itemsets. Hash tree needs two parameters (*child_max_size* and *leaf_max_size*) to be fine tuned for better performance and the same value of these parameters may not be suitable for different datasets and different minimum threshold [5].

## 2.3 Trie vs. Hash Table Trie

Support counting with a trie becomes slower when one has to move downward from a node having many links to the nodes at next lower level. There is a need to make a linear search at each node to move downward which results into an increased counting time. So to accelerate the search time we employ hash table, an efficient searching technique. Each node maintains a hash table and in this way we require steps just equal to the size of an itemset to reach to the leaf node. A perfect hashing have to be maintained since a leaf in a trie represents exactly one itemset. We named a trie with hashing technique as a hash table trie [6].

## 2.4 Apache Hadoop and MapReduce

The fundamental design principle of Hadoop is to distribute the computing power to where the data is. Moving data is much more costly than movement of computation as in well known parallel and distributed computing paradigm MPI (Message Passing Interface) data is being moved. Hadoop is extremely scalable and fault tolerant distributed system which minimizes the consumption of network bandwidth. It hides the parallelization, data distribution and load balancing [8-9]. Hadoop is as an open source project supported by Apache foundation [4] which is inspired by Google' File System (GFS) [10] and Google's MapReduce [11] programming model.

Two major components of Hadoop are Hadoop Distributed File System (HDFS) and Hadoop MapReduce. Hadoop processes data residing in HDFS using MapReduce. HDFS architecture is based on GFS. It is a highly scalable storage and supports fast accessing to large datasets. It stores and files by breaking them into blocks and replicates the blocks across multiple nodes to facilitate high availability and fault tolerance. Default block size is 64 MB and default replication factor is 3 [9].

MapReduce is a programming paradigm of Hadoop for parallel and distributed computation on large datasets, which is based on the underlying ideas of Google's MapReduce. To compute the problem using MapReduce one has to divide the computation into two tasks: a map and a reduce task. Input dataset is splitted into smaller chunks and assigned to mapper executing map task. Chunk size is customized by InputFormat class of MapReduce framework. Each chunk is assigned to an individual mapper. Mapper processes the assigned datasets and output a number of (key, value) pairs. MapReduce framework automatically sort and shuffle these (key, value) pairs for a number of Reducers to execute reduce task. In shuffling process a reducer is assigned key and list of values associated with that key. Reducer processes key and list of values and output new (key, value) pairs. An additional Combiner function may be used to reduce the data transfer from mappers to reducers. It works same as Reducer but only on (key, value) pairs generated by mappers on one node.

MapReduce framework allows a single time communication when outputs of mappers are transferred to reducers. All the mappers and reduces are executed independently without any communication between them [9].

# 3. RELATED WORKS

A number of Apriori based algorithms have been proposed to reduce the computation time and to enhance the scalability. Among sequential algorithm most significant are based on efficient data structures. Parallel and distributed algorithms are developed to improve the scalability as well as efficiency. But these traditional parallel and distributed algorithms are not efficient and scalable to manage big data. Hadoop provides an extremely scalable and fault tolerant cluster system so it is unavoidable to re-design existing data mining algorithms on MapReduce framework in order to execute them on Hadoop cluster.

The A number of MapReduce based Apriori algorithms have been proposed but most of them are simply straight forward implementation of Apriori [12-16]. Fixed Passes Combined-counting (FPC) and Dynamic Passes Combined-counting (DPC) [17] are the two algorithms which significantly reduces the execution. These algorithms are based on combining multiple consecutive passes of Single Pass Counting (SPC) algorithm [17] in a single map-reduce phase. SPC is a straight forward implementation of Apriori. Algorithm proposed by F. Kovacs and J. Illes [18] is most likely to be SPC except candidates are generated inside reducer as it is kept inside mapper traditionally in most of the algorithms. Authors also used triangular matrix data structure to count 1 and 2-itemsets in a single step. L. Li and M. Zhang [19] proposed a one phase map-reduce algorithm to generate frequent itemsets. It generates frequent 1 to k-itemsets in a single map-reduce phase. This algorithm does not strictly follow Apriori algorithm since using one map-reduce completely skips pruning steps based on Apriori property. Pruning steps reduces the set of candidates by checking against frequent itemsets of last iteration. But authors proposed a dataset distribution method for heterogeneous Hadoop cluster. Honglie Yu *et al.* [20] proposed an algorithm based on Boolean matrix and applied AND operation on it. In this algorithm transactional database is replaced by Boolean matrix and divide the matrix on various DataNodes of Hadoop cluster. The PARMA, a parallel randomized algorithm on MapReduce is proposed by Matteo Riondato *et al*. [21], which is independent from dataset size. It discovers approximate frequent itemsets from a small sample of datasets. A comprehensive and more descriptive literature review of MapReduce based Apriori algorithm can be found in [22].

In order to implement Apriori algorithm on MapReduce framework, role of data structures have not been evaluated in MapReduce context. Hash tree and trie are the central data structures for Apriori algorithm [2] [5]. F. Bodon and L. Rónyai [5] proposed the trie data structure and also proved theoretically and experimentally that trie is faster, consumes less memory and simpler to generate candidates in comparison to hash tree. Christian Borgelt [23] efficiently implemented Apriori algorithm using trie. He represented the transactions in a trie to reduce the support counting cost. He also proposed the idea of transaction filtering by removing infrequent items from transactions. F. Bodon proposed a number of strategies to improve the performance of sequential Apriori using trie data structure [24-25]. He proposed many routing strategies at the nodes, storing frequent and candidate itemsets in a single trie. F. Bodon and L. Schmidt-Thieme [26] proposed an efficient intersection-based pruning method which saves superfluous traversal of some part of trie. Trie with hashing technique has been proposed by F. Bodon [6] to speed up the search time in trie. Theoretically hashing technique in trie seems to accelerate the support counting but the experimental results have not shown any improvement. Author left this technique for further investigation.

# 4. IMPLEMENTATIONS

A computational problem is submitted as a MapReduce Job on Hadoop cluster. A job is configured using a Driver class. A driver class is defined with a Mapper class, a Reducer and an optional Combiner class of MapReduce framework. Also input/output directory in HDFS, input split size and other problem specific parameters are specified in driver class. To implement Apriori algorithm on MapReduce framework we have to split it into two independent sub-problems corresponding to map and reduce tasks. We define two sub-problems candidate itemsets generation and frequent itemsets generation and assigned them to Mapper and Reducer respectively. Each mapper processes a chunk of input datasets and generates local candidates with local support count. Each reducer receives local candidates and sums up the local count to generate frequent itemsets. Number of mappers depends on the number of chunks so if we reduce the chunk size there will be more mappers. Number of reducers is specified in driver class and does not depend on input size. All the mappers and reducers execute in parallel across different nodes of cluster but final result cannot be obtained until all reduce tasks are not completed. Apriori is an iterative algorithm so we have to submit job each time a new iteration starts.

Again we have to define two types of jobs, one for frequent 1-itemset generation and other for frequent k-itemset ($k \geq 2$) generation. In both type of jobs the functionality of combiner and reducer remains same since they only make sums up the local count. We define two mappers corresponding to 1-itemsets and k-itemsets. Algorithm 1 depicts the pseudo code for the driver class of the Apriori algorithm. Job1 is executed only once to generate frequent 1-itemsets and Job2 is executed iteratively to generate frequent k-itemsets, until further candidate generation are not possible. Here OneItemsetMapper generates candidate 1-itemsets and K-ItemsetMapper generates candidate k-itemsets. Pseudo code for OneItemsetMapper and K-ItemsetMapper are shown in Algorithm 2 and 3. ItemsetCombiner makes the local sum of the local candidates on one node. ItemsetReducer sums up the local count of the candidates obtained from all the nodes and check against minimum support threshold. Candidates satisfying minimum support threshold are produced as frequent itemsets. Algorithm 4 depicts the pseudo code for ItemsetCombiner and ItemsetReducer. Pseudo code of ItemsetCombiner and ItemsetReducer are same except latter one make use of minimum support threshold.

**Algorithm 1.** DriverApriori

```
// Find frequent 1-itemset L1
  Job1: //submitted single time
      OneItemsetMapper
      ItemsetCombiner
      ItemsetReducer
  end Job1
// Find frequent k-itemset Lk
  for (k = 2; Lk-1 ≠ ϕ; k++)
      Job2: // submitted multiple times
          K-ItemsetMapper
          ItemsetCombiner
```

```
        ItemsetReducer
    end Job2
end for
```

**Algorithm 2.** OneItemsetMapper, k = 1

*Input:* a block $b_i$ of database
*key: byte offset of the line,*
*value: a transaction* $t_i$
*for each* $t_i \in b_i$ *do*
    *for each item* $i \in t_i$ *do*
        *write (i, 1);*
    *end for*
*end for*

**Algorithm 3.** K-ItemsetMapper, $k \geq 2$

*Input:* a block $b_i$ of database and $L_{k-1}$
*key: byte offset of the line,*
*value: a transaction* $t_i$
*//* $L_{k-1}$ *may be a Hash Tree, Trie or Hash Table Trie*
*read (k-1)-itemsets from cache file in* $L_{k-1}$
*//* $C_k$ *may be a Hash Tree, Trie or Hash Table Trie*
$C_k$ = *apriori-gen*($L_{k-1}$);
*for each* $t_i \in$ *block* $b_i$ *do*
    $C_t$ = *subset*($C_k$ , $t_i$); *//* $C_t$ *may be a List*
    *for each candidate* $c \in C_t$ *do*
        *write (c, 1);*
    *end for*
*end for*

**Algorithm 4.** ItemsetCombiner and ItemsetReducer

***ItemsetCombiner***
*key: itemset,*
*value: key's value list*
*for each key k do*
    *for each value v of k's value list*
        *sum += v;*
    *end for*
    *write(k, sum)*
*end for*

***ItemsetReducer***
*key: itemset,*
*value: key's value list*
*for each key k do*
    *for each value v of k's value list*
        *sum += v;*
    *end for*
    *if sum >= min_supp_count*
        *write(k, sum)*
    *end if*
*end for*

Algorithm for K-ItemsetMapper is central to our discussion. We have implemented three variants of K-ItemsetMapper for hash tree, trie and hash table trie. The algorithm remains unchanged and we have simply changed the data structure each time as shown in Algorithm 3. Method apriori-gen() and subset() are the most computation intensive steps, which generate candidates and count support respectively using considered data structure. So operation cost of apriori-gen() and subset() methods are greatly affected by the data structure used.

We have implemented the three data structures in Java to be used in MapReduce code. To implement hash tree, we have defined two classes named as InnerNode and LeafNode for inner node and leaf node of hash tree. InnerNode contains a list of size *child_max_size*, which contains child nodes. These child nodes may be inner node or leaf node. LeafNode contains a list of size *child_max_size*, which again contains a list of candidates. Thus candidates are stored at leaf node. Candidate generation and support counting using hash tree are implemented following the techniques mentioned in original works.

All nodes in a trie have same structure. We have defined a class TrieNode for the node of trie. TrieNode contains a String object to store item label of a link, address of parent node and a list of child nodes. For a given itemset, we traverse downward in a trie by searching item label of links at each node. Again methods for candidate generation and support counting using trie are followed as it is in [5].Hash table trie is implemented using perfect hashing at each node. We just modified the class TrieNode of trie and added a hash table in it. Because of that we have used hashing technique instead of searching item label linearly at each node.

# 5. EXPERIMENTAL RESULTS

In this section we have observed the execution time of algorithms when it uses hash tree, trie and hash table trie for different minimum support value. Also execution times for different sizes of input split are also evaluated. We have analyzed the execution time on varying minimum support values and with increasing number of mappers.

## 5.1 Cluster Setup and Datasets

A small Hadoop-2.6.0 cluster is installed with five nodes, all are running Ubuntu 14.04. One node is dedicated as NameNode and other four nodes serve as DataNodes. NameNode is configured with 4 cores and 4 GB memory running in the virtualized environment on window host. Two DataNodes are running on separate physical machine each with 4 cores and 2 GB memory. Other two DataNodes with 4 cores and 4 GB memory are running in the virtualized environment on another same window host. All algorithms are implemented in Java and used MapReduce 2.0 library.

Experiments were carried out on both real life and synthetic datasets. Two real life datasets BMS_WebView_1 and BMS_WebView_2 are click-stream data from a web store used in KDD-Cup 2000 [27]. The synthetic dataset T10I4D100K is generated by IBM generator [28]. BMS_WebView_1 contains 59602 transactions with 497 items and BMS_WebView_2 contains 77512 transactions with 3340 items.

## 5.2 Execution Time for different Minimum Support Values

We have observed the execution time of algorithms on the above three datasets for different minimum support values. We have used 4 reducers in all algorithms. The number of mappers is dependent on the number of input split since for each split one mapper is assigned. As the chunk size i.e. the number of lines of input per split decreases, the number of splits increases. We have set 5K and 6.5K as chunk size for BMS_WebView_1 and BMS_WebView_2 respectively, which results 12 mappers. Chunk size for T10I4D100K dataset was 5K, which results 20 mappers. Figure 2, 3 and 4 shows the execution time on datasets BMS_WebView_1, BMS_WebView_2 and T10I4D100K respectively for varying value of minimum support. The performance of hash table trie is outstanding in comparison to trie and hash tree on all datasets. The performance of trie is far better than hash tree on BMS_WebView_1 dataset but it is not in the case of BMS_WebView_2. Also on T10I4D100K dataset, trie performs nearly same as hash tree.

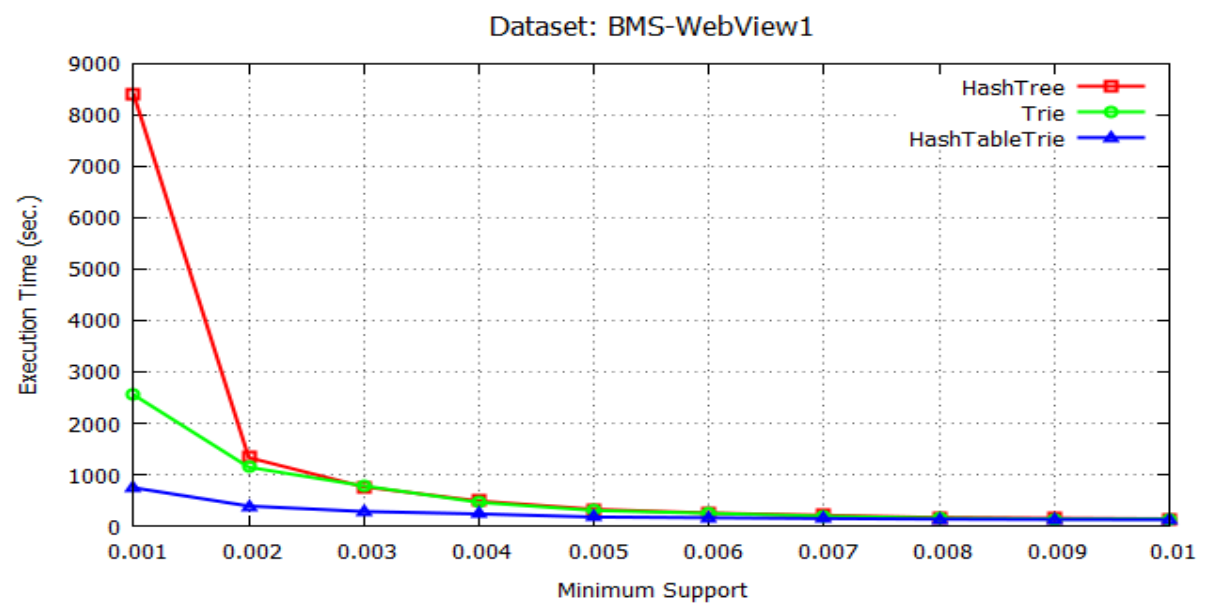

**Fig 2: Execution time of Apriori using three data structures on dataset BMS_WwbView_1**

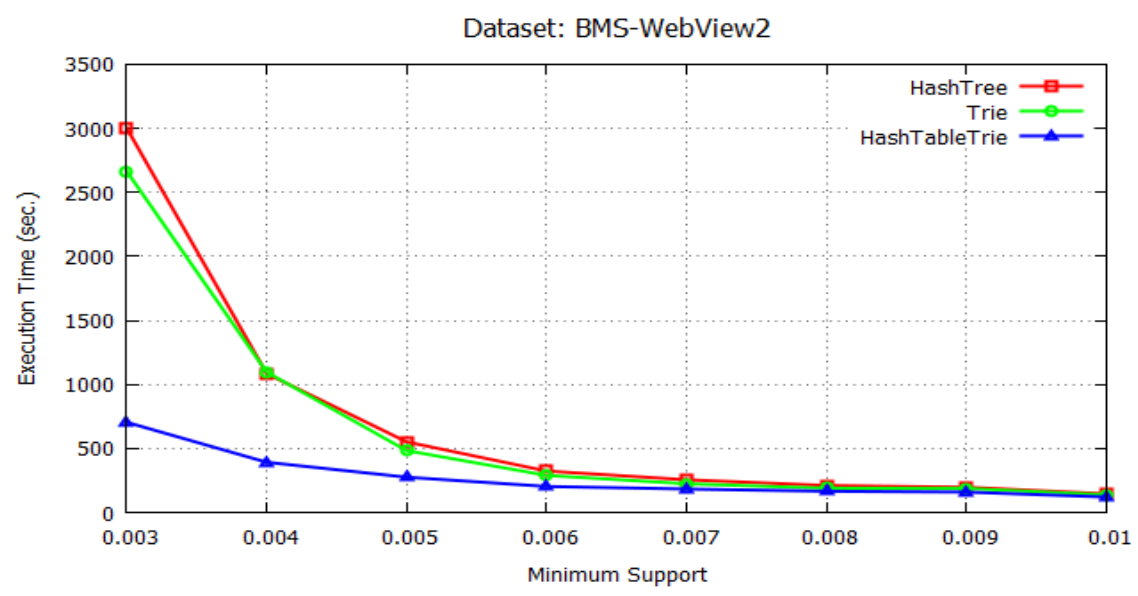

**Fig 3: Execution time of Apriori using three data structures on dataset BMS_WwbView_2**

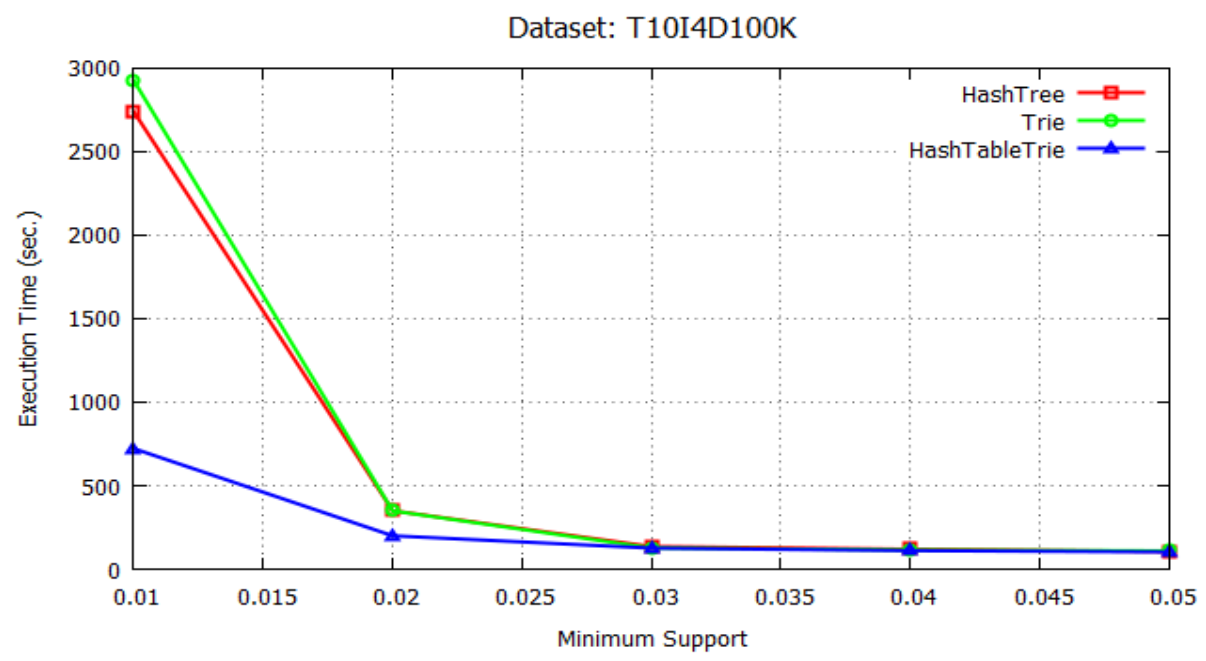

**Fig 4: Execution time of Apriori using three data structures on dataset T10I4D100K**

In hash tree we have only considered the parameter *child_max_size* with value 20. We have ignored the second parameter *leaf_max_size* for simplicity of implementation. As we have discussed in earlier section that performance of hash tree depends on these two parameters and its major downside is that value of parameters suitable for one datasets may not be for others. This can be observed form the Figure 2, 3 and 4 for three datasets. Hash tee performs worst on dataset BMS_WebView_1 while its performance is competing trie on BMS_WebView_2 and T10I4D100K. We have explored this case iteration-wise for dataset BMS_WebView_2. All algorithms made 7 iterations (jobs) for minimum support 0.003 on this dataset. Table 1 compares the execution time of these 7 iterations of hash tree and trie based algorithms. Here we observed that the performance of trie deteriorates when we generate 2-itemsets in iteration 2 whereas in other iterations it performs better than or equivalent to hash tree.

**Table 1. Execution time (sec.) of respective iterations for hash tree and trie**

| **Iteration No.** | 1 | 2 | 3 | 4 | 5 | 6 | 7 |
|---|---|---|---|---|---|---|---|
| **Hash Tree** | 23 | 2078 | 164 | 300 | 252 | 66 | 26 |
| **Trie** | 37 | 2432 | 39 | 38 | 32 | 28 | 23 |

The most important reason behind outstanding performance of hash table trie may be the memory it requires during execution. MapReduce starts a new job each time a next iteration of Apriori starts. In $k^{th}$ iteration there are only frequent (k-1)-itemsets and candidates k-itemsets reside in the memory and other itemsets of previous iterations are discarded. Reason for unsuccessfulness of hash table trie in sequential Apriori discussed by the author [6] is as follows. Using hash table enlarges the size of a node, which could not be cached in and may be moved into memory. Linear search is fast in cache and reading operation is slower for memory. Since in MapReduce iteration, there is lesser number of nodes in memory. Therefore, searching and reading operations will be faster for small number of nodes.

## 5.3 Execution Time for Increasing Number of Mappers

In MapReduce framework, degree of parallelism can be increased in two ways, either by increasing DataNodes or by increasing mappers. Number of DataNodes can be controlled by editing list of slaves in slave file on NameNode. Number of mappers can be controlled directly in MapReduce code using NLineInputFormatClass which set the chunk size (number of lines per split). A smaller chunk size results a larger number of mappers. Table 2 shows the execution times of three algorithms on dataset T10I4D100K with minimum support 0.02, for increasing number of mappers.

**Table 2. Execution time of Apriori using hash tree, trie and hash table trie for different chunk size**

| **Chunk Size** | **Number of Mappers** | **Execution Time (sec.)** | | |
|---|---|---|---|---|
| | | **Hash Tree** | **Trie** | **Hash Table Trie** |
| 100K | 1 | 2907 | 2892 | 1124 |
| 50K | 2 | 1649 | 1442 | 584 |
| 20K | 5 | 720 | 657 | 293 |
| 10K | 10 | 425 | 430 | 214 |
| 5K | 20 | 350 | 349 | 200 |

From Table 2 it is clear that as the number of mappers increases the execution time for all data structures decreases. But reduction in execution time also turns to lower with increasing number of mappers. For example, the difference in execution time for 10 and 20 mappers is not significant. Increasing the number of mappers does not work after a particular point since communication and scheduling overhead have been increased along with. We can also represent it in terms of the speedup using multiple mappers. Speedup is the ratio of execution time without improvement and improved execution time. We defined the speed up in our case as following.

*Speed up = Execution time with 1 mapper / Execution time with N mappers*

Figure 5 shows the speedup calculated from Table 2, of Apriori using hash tree, trie and hash table trie, for varying number of mappers. We can see that up to 10 mappers we have achieved a good speedup but after that increasing the number of mappers have shown no significant achievement in speedup.

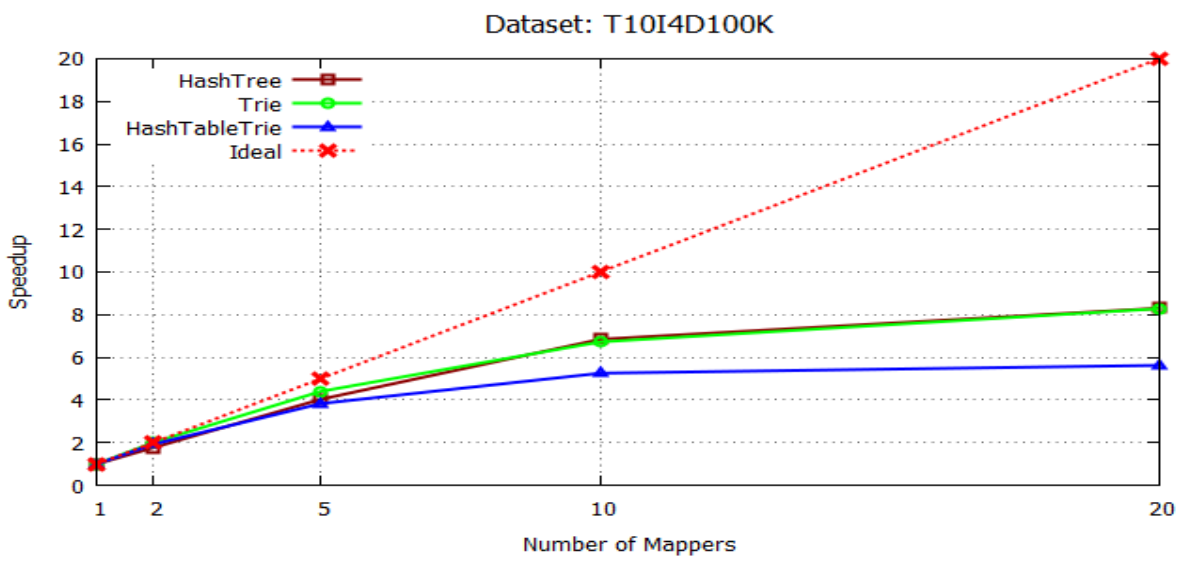

**Fig 5: Speedup for different number of Mappers**

## 6. CONCLUSIONS

Frequent itemset mining algorithms are the most researched field of data mining. Re-designing data mining algorithms on MapReduce framework to analyze big data is the new drift in research. We have identified and filled up the gap between effects of data structures on traditional Apriori and on MapReduce based Apriori. We have implemented the Apriori algorithm on MapReduce framework, for the central data structures hash tree, trie and hash table trie and evaluated the execution time of Apriori algorithms for these data structures in MapReduce context. In sequential computing environment, trie outperformed hash tree experimentally as well as theoretically. Hash table trie theoretically accelerates the Apriori algorithm but experimentally could not success. In our experiment on Hadoop cluster, trie outperforms hash tree as usual but hash table trie is outstanding among the three data structures for both real-life and synthetic datasets. Reason behind it may be the lesser number nodes residing in memory since in MapReduce itemsets of previous iterations are discarded. For small number of nodes, searching and reading operations will be faster. Further possible implementations can be checked in order to improve the performance of MapReduce based Apriori. One can implement the existing idea of using mixed of simple trie node and hash table trie node, and also deploying a joint trie for both frequent and candidate itemsets.

## 7. REFERENCES

[1] Agrawal, R., Imielinski, T. and Swami, A. 1993. Mining Association Rules between Sets of Items in Large Databases. In ACM SIGMOD Conf. Management of Data, Washington, D.C., 207–216.

[2] Agrawal, R. and Srikant, R. 1994. Fast Algorithms for Mining Association Rules. In Proceedings of the Twentieth International Conference on Very Large Databases, Santiago, Chile, 487–499.

[3] Ward, J. S. and Barker, A. Undefined By Data: A Survey of Big Data Definitions. http://arxiv.org/abs/1309.5821v1. Retrieved Sept. 2015.

[4] Apache Hadoop. http://hadoop.apache.org

[5] Bodon, F. and Rónyai, L. "Trie: an alternative data structure for data mining algorithms", Mathematical and Computer Modelling, 2003, 38(7), 739-751.

[6] Bodon, F. 2010. A fast apriori implementation. In Proceedings of IEEE ICDM workshop on frequent itemset mining implementations (FIMI'03), Vol. 90.

[7] Han, J. and Kamber, M. 2006. Data Mining: Concepts and Techniques. Morgan Kaufmann Publishers.

[8] HDFS Architecture Guide. https://hadoop.apache.org/docs/r1.2.1/hdfs_design.html

[9] Yahoo! Hadoop Tutorial. http://developer.yahoo.com/hadoop/tutorial/index.html

[10] Ghemawat, S., Gobioff, H. and Leung, S. "The Google File System", ACM SIGOPS Operating Systems Review, 2003, 37(5), 29–43.

[11] Dean, J. and Ghemawat, S. "MapReduce: Simplified Data Processing on Large Clusters", ACM Commun., 2008, vol. 51, 107–113.

[12] Li, J., Roy, P., Khan, S. U., Wang, L. and Bai, Y. "Data Mining Using Clouds: An Experimental Implementation of Apriori over MapReduce", http://sameekhan.org/pub/L−K−2012−SCALCOM.pdf, Retrieved March 2014.

[13] Lin, X. 2014. MR-Apriori: Association Rules Algorithm Based on MapReduce. In Proceedings of IEEE International Conference on Software Engineering and Service Science (ICSESS).

[14] Yang, X. Y., Liu, Z. and Fu, Y. 2010. MapReduce as a Programming Model for Association Rules Algorithm on Hadoop. In Proceedings of 3rd International Conference on Information Sciences and Interaction Sciences (ICIS), 99(102), 23–25.

[15] Li, N., Zeng, L., He, Q. and Shi, Z. 2012. Parallel Implementation of Apriori Algorithm based on MapReduce. In Proceedings of 13th ACIS IEEE International Conference on Software Engineering, Artificial Intelligence, Networking and Parallel & Distributed Computing, 236–241.

[16] Oruganti, S., Ding, Q. and Tabrizi, N. 2013. Exploring HADOOP as a Platform for Distributed Association Rule Mining. In FUTURE COMPUTING 2013 the Fifth International Conference on Future Computational Technologies and Applications, 62–67.

[17] Lin, M-Y., Lee, P-Y. and Hsueh, S-C. 2012. Apriori-based Frequent Itemset Mining Algorithms on MapReduce. In Proceedings of 6th International Conference on Ubiquitous Information Management and Communication (ICUIMC '12), ACM, New York, Article 76.

[18] Kovacs, F. and Illes, J. 2013. Frequent Itemset Mining on Hadoop. In Proceedings of IEEE 9th International Conference on Computational Cybernetics (ICCC), Hungry, 241–245.

[19] Li, L. and Zhang, M. 2011. The Strategy of Mining Association Rule Based on Cloud Computing. In Proceedings of IEEE International Conference on Business Computing and Global Informatization (BCGIN), 29–31.

[20] Yu, H., Wen, J., Wang, H. and Jun, L. An improved Apriori Algorithm Based on the Boolean Matrix and Hadoop", Procedia Engineering 15 (2011), Elsevier, 1827-1831.

[21] Riondato, M., DeBrabant, J. A., Fonseca, R. and Upfal, E. 2012. PARMA: A Parallel Randomized Algorithm for Approximate Association Rules Mining in MapReduce.

In Proceedings of the 21st ACM international conference on Information and Knowledge Management, 85-94.

[22] Singh, S., Garg, R. and Mishra, P. K. 2014. Review of Apriori Based Algorithms on MapReduce Framework. In Proceedings of International Conference on Communication and Computing (ICC - 2014), Elsevier Science and Technology Publications, 593–604.

[23] Borgelt, C. 2003. Efficient implementations of apriori and éclat. In Proceedings of IEEE ICDM workshop on frequent itemset mining implementations (FIMI'03).

[24] Bodon, F. 2004. Surprising Results of Trie-based FIM Algorithms. FIMI 2004.

[25] Bodon, F. 2005. A trie-based APRIORI implementation for mining frequent item sequences. In Proceedings 1st international workshop on open source data mining: frequent pattern mining implementations, ACM.

[26] Bodon, F. and Schmidt-Thieme, L. 2005. The relation of closed itemset mining, complete pruning strategies and item ordering in apriori-based fim algorithms. In Knowledge Discovery in Databases: PKDD, Springer Berlin Heidelberg, 437-444.

[27] SPMF Datasets. http://www.philippe-fournier-viger.com/spmf/index.php?link=datasets.php

[28] Frequent Itemset Mining Dataset Repository. http://fimi.ua.ac.be/data/